\documentclass[12pt]{article}
\usepackage{amsmath,amssymb,amsfonts,epsfig}
\usepackage[active]{srcltx}
\usepackage{graphics}
\usepackage{amssymb,amsmath}
\usepackage{cite}
\usepackage[usenames]{color}
\usepackage[usenames,dvipsnames]{pstricks}
\usepackage{epsfig}
\usepackage{pst-grad} 
\usepackage{pst-plot} 
\newcommand{\ct}{\cite}
\newcommand{\lb}{\label}

\newcommand{\bc}{\begin{center}}
\newcommand{\ec}{\end{center}}
\newcommand{\bd}{\begin{displaymath}}
\newcommand{\ed}{\end{displaymath}}
\newcommand{\be}{\begin{equation}}
\newcommand{\ee}{\end{equation}}
\newcommand{\ba}{\begin{array}}
\newcommand{\ea}{\end{array}}
\newcommand{\bea}{\begin{eqnarray}}
\newcommand{\eea}{\end{eqnarray}}
\newcommand{\bt}{\begin{tabular}}
\newcommand{\et}{\end{tabular}}
\newcommand{\un}{\underline}

\newcommand{\bp}{\begin{picture}}
\newcommand{\ep}{\end{picture}}
\newcommand{\bfi}{\begin{figure}}
\newcommand{\efi}{\end{figure}}

\begin{document}

\begin{titlepage}

\begin{center}
{{\Large {\bf Baryogenesis in Cosmological Model \\ with
Superstring-Inspired $\Large \bf E_6$ Unification} }}
\vspace*{10mm}


{\bf \large{ C. R. Das $^{a}$,  L. V. Laperashvili $^{b}$, H.
B.~Nielsen $^{c}$ and A. Tureanu $^{d}$}}
\end{center}
\begin{center}
\vspace*{0.2cm} {\it { $ ^a$\large Centre for Theoretical Particle
Physics, Technical University of Lisbon, \\ Avenida Rovisco Pais,
1 1049-001 Lisbon,
Portugal\\
$^b$ The Institute of Theoretical and Experimental Physics,\\
Bolshaya Cheremushkinskaya, 25, 117218 Moscow, Russia \\
$^c$ Niels Bohr Institute, Blegdamsvej 17-21, DK-2100 Copenhagen, Denmark \\
$^d$ Department of Physics, University of Helsinki\\ and Helsinki
Institute of Physics, P.O.Box 64, FIN-00014 Helsinki, Finland}}

\footnote{crdas@cftp.ist.utl.pt, laper@itep.ru, hbech@nbi.dk,
anca.tureanu@helsinki.fi}

\end{center}

\begin{center}{\bf Abstract}\end{center}
\begin{quote}

We have developed a concept of parallel existence of the ordinary
(O) and  hidden (H) worlds with a superstring-inspired $E_6$
unification, broken at the early stage of the Universe into
$SO(10)\times U(1)$ -- in the O-world, and $SU(6)'\times SU(2)'$
-- in the H-world. As a result, we have obtained in the hidden
world the low energy symmetry group $G'_{SM}\times
SU(2)'_{\theta}$, instead of the Standard Model group $G_{SM}$.
The additional non-Abelian $SU(2)'_{\theta}$ group with massless
gauge fields, ''thetons", is responsible for the dark energy. We
present a baryogenesis mechanism with the $B-L$ asymmetry produced
by the conversion of ordinary leptons into particles of the hidden
sector.

\end{quote}

\end{titlepage}
\vspace*{0.2cm}

\section{Introduction}

A cosmological model has been proposed in Ref. \ct{1} with the
superstring-inspired $E_6$ unification arising at the early stage
of the Universe. Considering a parallel existence of the ordinary
(O) and hidden (H) worlds, it was assumed that the $E_6$ group was
broken differently in the O- and H-sectors with the following
breakings:
\be E_6\to SO(10)\times U(1) \lb{1A} \ee
-- in the O-world, and
\be E_6' \to SU(6)'\times SU(2)'  \lb{2A} \ee
-- in the H-world \footnote{The superscript 'prime' denotes the
H-world.}.

Using the model \ct{1}, we have tried to explain the origin of the
Dark Energy (DE), Dark Matter (DM) and visible matter with energy
densities given by recent cosmological observations, confirming
the $\Lambda CDM$ cosmological model with a tiny value of the
cosmological constant. The study \ct{1} is a development of the
ideas considered previously in Refs.\ct{1a}. In the present
investigation we describe the inflation epoch of our Universe and
baryogenesis scenario.

For the present epoch, the Hubble parameter $H=H_0$ is given by
the following value \ct{2,3}:
\be H_0 = 1.5 \times 10^{-42}\,\,{\rm{GeV}} \lb{1B}  
\ee
and the critical density of the Universe is
\be \rho_{c} = 3H^2/8\pi G = {(2.5\times 10^{-12}\,\,
{\rm{GeV}})}^4. \lb{2B} 
\ee
Cosmological measurements give the following density ratios of the
total Universe \ct{2,3}:
\be \Omega = \Omega_r + \Omega_m + \Omega_\Lambda = 1,
                  \lb{3B}  
\ee
where $\Omega_r \ll 1$ is a relativistic (radiation) density ratio
and
\be \Omega_{\Lambda} = \Omega_{DE}\sim 75\%\, \lb{4B}  
\ee
for the mysterious Dark Energy (DE), which is responsible for the
accelerated expansion of the Universe. The matter density ratio
is:
\be
 \Omega_m\approx \Omega_M + \Omega_{DM} \sim 25\%, \lb{5B} 
\ee
with $\Omega_M \approx \Omega_B \approx 4\%$ -- for (visible)
baryons, and $\Omega_{DM} \approx 21\% $ -- for the Dark Matter
(DM).
We can calculate the dark energy density using (\ref{2B}) and
(\ref{4B}):
\be \rho_{DE}= \rho_{vac} \approx 0.75\ \rho_c \approx (2.3 \times
10^{-3} \,\, {\rm eV})^4. \lb{6B} \ee
The result (\ref{6B}) is consistent with the present model of
accelerating Universe dominated by a tiny cosmological constant
and Cold Dark Matter (CDM).

\section{Superstring theory and $E_6$ unification}

Superstring theory \ct{6,7,8} is a paramount candidate for the
unification of all fundamental gauge interactions with gravity.
The 'heterotic' superstring theory $E_8\times E'_8$ reasonably was
suggested as a realistic model for unification of all fundamental
gauge interactions with gravity \ct{7}. This ten-dimensional
theory can undergo spontaneous compactification. The integration
over six compactified dimensions of the $E_8$ superstring theory
leads to the effective theory with the $E_6$ unification in the
four-dimensional space \ct{8}.

Superstring theory has led to the speculation that there may exist
another form of matter -- hidden ``shadow matter'' -- in the
Universe, which only interacts with ordinary matter via gravity or
gravitational-strength interactions \ct{9} (see also the reviews
\ct{9a}). The shadow world, in contrast to the mirror world
\ct{10}, can be described by another group of symmetry (or by a
chain of groups of symmetry), which is different from the ordinary
world symmetry group.

Three 27-plets of $E_6$ contain three families of quarks and
leptons, including right-handed neutrinos $N_i^c$ (where $i=1,2,3$
is the index of generations). We omit generation subscripts, for
simplification.

Matter fields (quarks, leptons and scalar fields) of the
fundamental 27-representation of the $E_6$ group decompose under
$SU(5)\times U(1)_X$ subgroup as follows (see Ref.~\ct{11}):
\be
        27 \to (10,1)  + (\bar 5, 2)+
          (5,-2)+ (\bar 5,-3)  + (1,5) + (1,0).    \lb{1a} \ee
The first and second numbers in the brackets in Eq. (\ref{1a})
correspond to the dimensions of the $SU(5)$ representations and to
the $U(1)_X$ charges, respectively. These representations decompose
under the groups with the breaking
\be SU(5)\times U(1)_X \to SU(3)_C\times SU(2)_L\times
U(1)_Z\times U(1)_X.            \lb{3a} \ee
We consider the following $U(1)_Z\times  U(1)_X$ charges of matter
fields: $Z=\sqrt{\frac{5}{3}}Q^Z,\,X=\sqrt{40}Q^X$.

The Standard Model (SM) family which contains the doublets of
left-handed quarks $Q$ and leptons $L$, right-handed up and down
quarks $u^c$, $d^c$, and also right-handed charged lepton $e^c$,
belongs to the $(10,1) + (\bar 5,2)$ representations of
$SU(5)\times U(1)_X$. Then, for the decomposition (\ref{3a}), we
have the following assignments of particles:
\bea
       (10,1) \to Q = &\left(\begin{array}{c}u\\
                                          d \end{array}\right) &\sim
                         \left(3,2,\frac 16,1\right),\nonumber\\
&u^{\rm\bf c} &\sim \left(\bar3,1,-\frac 23,1\right),\nonumber\\
&e^{\rm\bf c} &\sim \left(1,1,1,1\right).       \lb{4a}\\
(\bar 5,2) \to &d^{\rm\bf c}&\sim \left(\bar 3,1,\frac
13,2\right),
\nonumber\\
L = &\left(\begin{array}{c}e\\
                                             \nu \end{array}\right) &\sim
                         \left(1,2,-\frac 12,2\right),               \lb{5a}\\
(1,5) \to & S\,\,\, &\sim \,\,\left(1,1,0,5\right).\lb{6a} \eea
The remaining representations in (\ref{3a}) decompose as follows:
\bea
        (5,-2) \to& D&\sim \left(3,1,-\frac 13,-2\right),\nonumber\\
                   h = &\left(\begin{array}{c}h^+\\
                                               h^0 \end{array}\right) &\sim
                         \left(1,2,\frac 12,-2\right).
                                                              \lb{7a}\\
    (\bar 5,-3) \to &D^{\rm\bf c} &\sim \left(\bar 3,1,\frac 13,-3\right),
\nonumber\\
                     h^{\rm\bf c} = &\left(\begin{array}{c}h^0\\
                                               h^- \end{array}\right) &\sim
                         \left(1,2,-\frac 12,-3\right).              \lb{8a}
\eea
To the representation (1,5) is assigned the SM-singlet field S,
which carries nonzero $U(1)_X$ charge. The light Higgs doublets are
accompanied by the heavy colour triplets of exotic quarks
('diquarks') $D,D^{\rm\bf c}$ which are absent in the SM (see
Ref.~\ct{11}).

The right-handed heavy neutrino is a singlet field $N^c$
represented by (1,0):
\be
       (1,0) \to N^{\rm\bf c} \sim  (1,1,0,0).             \lb{9a}
\ee

\section{Breaking of the $E_6$ unification in cosmology}

The results of Refs.~\ct{12} are
based on the hypothesis of the existence in Nature a mirror (M)
world parallel to the visible ordinary (O) world. The authors have
described the O- and M-worlds at low energies by a minimal
symmetry $G_{SM}\times G'_{SM}$ where
$$G_{SM} = SU(3)_C\times SU(2)_L\times U(1)_Y$$ stands
for the observable Standard Model (SM) while
$$G'_{SM} = SU(3)'_C\times SU(2)'_L\times U(1)'_Y$$ is its mirror
gauge counterpart. The M-particles are singlets of $G_{SM}$ and
the O-particles are singlets of $G'_{SM}$. These different O- and
M-worlds are coupled only by gravity, or possibly by another very
weak interaction.

If the ordinary and mirror worlds are identical, then O- and
M-particles should have the same cosmological densities. But this
is immediately in conflict with recent astrophysical
measurements. Mirror parity (MP) is not conserved, and the
ordinary and mirror worlds are not identical. Then the VEVs of the
Higgs doublets $\phi$ and $\phi'$ are not equal:
\be \langle\phi\rangle=v,\quad \langle\phi'\rangle=v'\quad
{\rm{and}} \quad v\neq v'.  \lb{1} \ee
Introducing the parameter characterizing the violation of MP:
\be       \zeta = \frac{v'}{v} \gg 1, \lb{1b}  \ee
we have the estimate of Refs.~\ct{12}: $\zeta \sim 100$. Then the
masses of fermions and massive bosons in the mirror world are
scaled up by the factor $\zeta$ with respect to the masses of
their counterparts in the ordinary world:
\be
               m'_{q',l'}=\zeta m_{q,l}, \qquad
                 M'_{W',Z',\Phi'}=\zeta M_{W,Z,\Phi}, \lb{3b}
                 \ee
while photons and gluons remain massless in both worlds.

In contrast to Refs.~\ct{12}, in the present paper we consider a
cosmological model with $E_6$ unification when at the early stage
of the Universe the O- and H(exactly M)- worlds have the same
GUT-scales and GUT-coupling constants: $M_{E6}=M'_{E6'}$ and
$g_{E6}=g'_{E6'}$. Later the $E_6$ unification undergoes the
breakdown which is different for O- and H-worlds.

It is well known (see \ct{14}) that there exist the following
three schemes for breaking of the $E_6$ group:
\begin{eqnarray}
i)\,\,\, E_6 &\to& SU(3)_1\times SU(3)_2\times SU(3)_3,\label{break1}\\
ii)\,\,\, E_6&\to& SO(10)\times U(1),\label{break2}\\
iii)\,\,\, E_6&\to& SU(6)\times SU(2)\label{break3}.
\end{eqnarray}
The first case was considered in Ref.~\ct{1}, where the possibility
of the breaking
\be E_6\to SU(3)_C\times SU(3)_L\times SU(3)_R \lb{3} \ee
was investigated in the ordinary and mirror worlds, assuming broken
mirror parity. The model has the merit of an attractive simplicity.
However, in such a model one is unable to explain the tiny value of
cosmological constant given by astrophysical measurements, because
in the case (\ref{3})\, we have in both worlds the low-energy limit
of the Standard Model (SM), which forbids a large confinement radius
(i.e. small energy scale) of any interaction.

It is impossible to obtain the same $E_6$ unification in the O- and
M-worlds with the same breakings $ii)$ or $iii)$ if the mirror
parity is broken in the Universe. In this case, we are forced to
assume different breakings, (\ref{1A}) and (\ref{2A}), of the $E_6$
unification in the O- and H-worlds. Since astrophysical measurements
confirm zero contributions to the dark energy from both SM and SM'
sectors, we explain the small value of the cosmological constant
$\Lambda=\rho_{vac}=\rho_{DE}$ by condensation of fields belonging
to the additional $SU(2)'$ gauge group which exists only in the
H-world and has a large confinement radius.

The breaking mechanism of the $E_6$ unification is given in
Ref.~\ct{15}. The vacuum expectation values (VEVs) of the Higgs
fields $H_{27}$ and $H_{351}$ belonging to 27- and 351-plets of
the $E_6$ group can appear in the case (\ref{1A}) only with
nonzero 27-component:
\be  \langle H_{351}\rangle=0, \qquad v=\langle H_{27}\rangle\neq
0. \lb{4b} \ee
In the case (\ref{2A}) we have
\be  \langle H_{27}\rangle=0, \qquad V=\langle H_{351}\rangle\neq
0. \lb{5b} \ee
The 27 representation of $E_6$ is decomposed into 1 + 16 + 10
under the $SO(10)$ subgroup and the 27 Higgs field $H_{27}$ is
expressed in 'vector' notation as
\bea
   H_{27}\equiv
    &\left(\begin{array}{c}H_0\\H_{\alpha}\\
                                          H_M\end{array}\right),&
\eea
where the subscripts 0, $\alpha=1,2,...,16$ and $M=1,2,...,10$ stand
for  singlet, the 16- and the 10-representations of $SO(10)$,
respectively. Then
\bea
   \langle H_{27}\rangle=
    &\left(\begin{array}{c}v\\0\\
                                          0\end{array}\right).&
\eea
Taking into account that the 351-plet of $E_6$ is constructed from
$27\times 27$ symmetrically, we see that the trace part of
$H_{351}$ is a singlet under the maximal little groups. Therefore,
in a suitable basis, we can construct the VEV $\langle
H_{351}\rangle$  for the case of the maximal little group
$SU(2)\times SU(6)$. A singlet under this group which we get from
a symmetric product of $27\times 27$ comes from the component $(1,
15)\times (1, 15)$ and hence
\bea \langle H_{351}\rangle =&\left(\begin{array}{cc}V\otimes
1_{15}&\\&0\otimes 1_{15}\end{array}\right).& \eea
According to the assumptions of Ref.~\ct{1}, in the ordinary
world, from the Standard Model (SM) scale up to the $E_6$
unification, there exists the following chain of symmetry groups:
$$SU(3)_C\times SU(2)_L\times U(1)_Y
\to  [SU(3)_C\times SU(2)_L\times U(1)_Y]_{{SUSY}}
$$ $$\to
 SU(3)_C\times SU(2)_L \times SU(2)_R\times U(1)_X\times
 U(1)_Z\to SU(4)_C\times SU(2)_L \times SU(2)_R\times U(1)_Z $$ \be \to
 SO(10)\times U(1)_Z \to E_6. \lb{6b}                
 \ee
In the shadow H-world, we have the following chain of symmetry
groups :
$$
SU(3)'_C\times SU(2)'_L\times SU(2)'_{\theta}\times U(1)'_Y \to
[SU(3)'_C\times SU(2)'_L\times SU(2)'_{\theta}\times
U(1)'_Y]_{{SUSY}}
$$ $$\to SU(3)'_C\times
SU(2)'_L\times SU(2)'_{\theta}\times U(1)'_X \times U(1)'_Z\to
SU(4)'_C\times SU(2)'_L\times SU(2)'_{\theta}\times U(1)'_Z $$ \be
\to SU(6)'\times SU(2)'_{\theta}\to E'_6.    \lb{7b} \ee
In general, this is not an unambiguous choice of the $E_6(E'_6)$
breaking chains.

\section{Shadow theta particles}

In the present paper we assume the
existence of the shadow low-energy symmetry group:
\be G' = SU(3)'_C\times SU(2)'_L\times SU(2)'_{\theta}\times
U(1)'_Y \,, \lb{8b} \ee
with an additional non-Abelian $SU(2)'_{\theta}$ group whose gauge
fields are neutral, massless vector particles -- thetons (see
Ref.~\ct{16}). This is a natural consequence of different schemes of
the $E_6$-breaking in the O- and H-worlds.  By analogy with the
theory developed in \ct{16}, we consider shadow thetons
${\Theta'}^i_{\mu\nu}$, $i=1,2,3$, which belong to the adjoint
representation of $SU(2)'_{\theta}$, three generations of shadow
theta-quarks $q'_{\theta}$ and shadow leptons $l'_{\theta}$, and the
necessary theta-scalars $\phi'_{\theta}$ for the corresponding
breakings.  The theta-particles are absent in the ordinary world
(they are not confirmed by experiment), however, they can exist in
the hidden world. We assume that shadow thetons have the macroscopic
confinement radius $1/\Lambda'_{\theta}$, where
$\Lambda'_{\theta}\sim 10^{-3}\,\, {\mbox{eV}}$.

\section{Inflation, $\bf E_6$ unification and the problem of walls in the Universe}

The simplest model of inflation is based on the
superpotential
\be W=\lambda \varphi(\Phi^2-\mu^2), \lb{1i} \ee
containing the inflaton field given by $\varphi$ and the Higgs field
$\Phi$, where $\lambda$ is a coupling constant of order 1 and $\mu$
is a dimensional parameter of the order of the GUT scale. The
supersymmetric vacuum is located at $\varphi=0$, $\Phi=\mu$, while
for the field values $\Phi=0$, $|\varphi| > \mu$ the tree level
potential has a flat valley with the energy density
$V=\lambda^2\mu^4$. When the supersymmetry is broken by the
non-vanishing F-term, the flat direction is lifted by radiative
corrections and the inflaton potential acquires a slope appropriate
for the slow roll conditions.

This so-called hybrid inflation model leads to the choice of the
initial conditions \ct{17}. Namely, at the end of the Planck epoch
the singlet scalar field $\varphi$ should have an initial value
$\varphi=f\sim 10^{18}$ GeV ($E_6$-GUT scale), while the field
$\Phi$ must be zero with high accuracy over a region much larger
than the initial horizon size $\sim M_{Pl}$. In other words, the
initial field configuration should be located right on the bottom of
the inflaton valley and the energy density starts with
$V=\lambda^2\mu^4 \ll M_{Pl}^4$.

If $E_6'$ is the mirror counterpart of $E_6$, then we have $Z_2$
symmetry, i.e. a discrete group connected with the mirror parity.
In general, the spontaneous breaking of a discrete group leads to
phenomenologically unacceptable walls of huge energy per area.
Then we have the following properties for the energy densities of
radiation, DM, M and wall:
$$ \rho_r\varpropto \frac 1{a(t)^4}, \quad \rho_{M,DM}\varpropto
\frac 1{a(t)^3}, \quad \rho_{wall}\varpropto \frac 1{a(t)}, $$
where $a(t)$ is a scale factor with cosmic time $t$ in the
Friedmann-Lemaitre-Robertson-Walker (FLRW) metric describing our
Universe. For large Universe we have
$\rho_{wall}\gg\rho_{M,DM},\rho_r.$ In our case of the hidden
world, the shadow superpotential is:
\be W'=\lambda' \varphi'({\Phi'}^2-\mu'^2), \lb{2i} \ee
where $\Phi'=H_{351}$ and $\langle H_{351}\rangle=\mu'$. Then the
initial energy density in the H-world is $V'={\lambda'}^2{\mu'}^4
\ll M_{Pl}^4$. To avoid this phenomenologically unacceptable wall
dominance we cannot assume symmetry under $Z_2$ and thus $V=V'$ is
not automatic. Instead, it is necessary to assume the following
fine-tuning:
\be V=V': \quad \lambda^2\mu^4={\lambda'}^2{\mu'}^4, \lb{3i} \ee
which helps to obtain the initial conditions for the GUT-scales
and GUT-coupling constants: $M_{E6}=M'_{E6'}$ and
$g_{E6}=g'_{E6'}$.

\section{Quintessence model of cosmology. Inflaton and axion}

Quintessence is described by a complex scalar field $\varphi$
minimally coupled to gravity.

We assume that there exists an axial $U(1)_A$ global symmetry in
our theory, which is spontaneously broken at the scale $f$ by a
singlet complex scalar field $\varphi$:
\be \varphi = (f + \sigma) \exp(ia_{ax}/f). \lb{1q} \ee
We assume that a VEV $\langle \varphi \rangle = f$ is of the order of
the $E_6$ unification scale: $f\sim 10^{18}$ GeV. The real part
$\sigma $ of the field $\varphi$ is the inflaton, while the boson
$a_{ax}$ (imaginary part of the singlet scalar fields $\varphi$)
is an axion and could be identified with the massless Nambu-Goldstone
(NG) boson if the corresponding $U(1)_A$ symmetry is not
explicitly broken by the gauge anomaly. However, in the hidden
world the explicit breaking of the global $U(1)_A$ by
$SU(2)'_\theta$ instantons inverts $a_{ax}$ into a pseudo
Nambu-Goldstone (PNG) boson $a_{\theta}$. Therefore, in the
H-world we have:
\be \varphi' = (f + \sigma') \exp(ia_{\theta}/f). \lb{2q} \ee
In Ref.~\ct{1} we have constructed a quintessence model of
cosmology with the axion $a_{\theta}$, having the mass $m\sim
{\Lambda'_{\theta}}^2/f\sim 10^{-42} \,\,{\rm{GeV}}$. Also we have
calculated the dark energy density due to the condensation of the
theta-fields:
\be \rho_{DE} = \rho_{vac}={(\Lambda'_\theta)}^4 \approx (2.3
\times 10^{-3} \,\, {\rm eV})^4.   \lb{3q} \ee
That is to say that {\em provided there were no other
contributions}, (\ref{3q}) would be our prediction for the
cosmological constant. It is the interesting point that this value
agrees very well with the phenomenological value (\ref{6B}).

The inflaton field provided the mechanism of rapid expansion after
the initial expansion that formed the Universe. Any inflationary
model has to describe how the SM-particles were generated at the
end of inflation. The inflaton, which is a singlet of $E_6$, can
decay, and the subsequent thermalization of the decay products can
generate the SM-particles. The inflaton $\sigma$ produces gauge
bosons: photons, gluons, $W^{\pm}$, $Z$, and matter fields:
quarks, leptons and the Higgs bosons, while the inflaton field
$\sigma'$ produces H-world particles: shadow photons and gluons,
thetons, $W'$, $Z'$, theta-quarks $q_{\theta}$, theta-leptons
$l_{\theta}$, shadow quarks $q'$ and leptons $l'$, scalar bosons
$\phi_{\theta}$ and shadow Higgs fields $\phi'$.

 In the shadow world we end up with a thermal bath of $SM'$ and $\theta$
 particles. However, as it was mentioned above, we assume that the density
 of $\theta$ particles is not too essential in cosmological
 evolution due to small $\theta$ coupling constants.

In the present model and in Refs.~\ct{18,19}, at the end of
inflation the O- and H-sectors are reheated in a non-symmetric way
($T_R > T'_R$). The reheating temperature $T_R$, at which the
inflaton decay and entropy production of the Universe are over,
plays a crucial role in cosmological evolution. In our model,
after the postinflationary reheating, the shadow sector is cooler
than the ordinary one and almost "empty". Therefore, the cosmology
of the early H-world is very different from the ordinary one when
we consider such crucial epochs as baryogenesis and
nucleosynthesis. Any of these epochs is related to an instant when
the rate of the relevant particle process, $\Gamma(T)$, becomes
equal to the Hubble expansion rate $H(T)$. In the H-world these
events take place earlier and the processes freeze out at larger
$T$ than in the ordinary world.

\section{Baryogenesis}

There is currently insufficient evidence to explain
why the Universe contains far more baryons than anti-baryons. The
first explanation for this phenomenon was given by A.D.~Sakharov
\ct{20}. The standard mechanism of baryogenesis is based on the
following three Sakharov conditions: 1. \un{B-violation}, which
was confirmed by cosmological inflation \ct{21}; 2. Breaking of
symmetry between particles and antiparticles, i.e. \un{C and
CP-violation}; 3. \un{Deviation from thermal equilibrium}.

Neither of these three conditions is obligatory. A lot of models
can explain the single observed number:
\be \beta_{observed}= \frac{n_B-n_{\bar B}}{n_{\gamma}}\approx
6\cdot 10^{-10}, \lb{1C} \ee
where $n_B,\,n_{\bar B},\,n_{\gamma}$ are baryon, anti-baryon and
$\gamma$ densities, respectively.

In our model, after the non-symmetric reheating with $T_R
> T'_R$, the exchange processes between O- and H-worlds are too
slow, by reason of the very weak interaction between the two
sectors. As a result, it is impossible to establish equilibrium
between them, so that both worlds evolve adiabatically and the
temperature asymmetry ($T'/T < 1$) is approximately constant in all
epochs from the end of the inflation until the present epoch.

The equilibrium between two sectors of massless particles with the
same temperature is not broken by the cosmological expansion, and
the baryon asymmetry (and any charge asymmetry) cannot be generated
in the Universe. However, if there are two components in the plasma
with different temperatures, then the equilibrium is explicitly
broken as long as the temperatures are not equal. In our case of
observed and hidden sectors, the equilibrium never happens by reason
of their essentially different temperatures. In this case, baryon
asymmetry may be generated even by scattering of massless particles.

Indeed, due to CP violation, the following cross-sections (with
ordinary quarks $q$ and hidden quarks $q'$):
\be   \sigma(q+q\to q'+q')\neq  \sigma(\bar q + \bar q\to \bar
{q'} + \bar {q'}) \ee
are different from each other. If we neglect the inverse process,
an asymmetry would be generated. The inverse process can be
neglected, being less efficient than the direct one, if the
temperature of the hidden sector is much lower than that of the
observed sector. So the baryon asymmetry can be generated even in
reactions with massless particles.

In the Bento-Berezhiani model of baryogenesis \ct{18} the heavy
Majorana neutrinos play the role of messengers between ordinary
and mirror worlds. Their model considers the group of symmetry
$G_{SM}\times G_{SM'}$, i.e. the Standard model and its mirror
counterpart. Heavy Majorana neutrinos $N$ are singlets of $G_{SM}$
and $G_{SM'}$ and this is an explanation, why they can be
messengers between ordinary and mirror worlds.

In our model with $E_6$ unification, the $N$-neutrinos belong to the
27-plet of $E_6$ and $E'_6$, and they are not singlet particles.
But after the breaking
\be E_6\to
 SO(10)\times U(1)_Z\to
 SU(3)_C\times SU(2)_L \times SU(2)_R\times U(1)_X\times
 U(1)_Z  \lb{2C} \ee
in the O-world, and
\be E'_6\to \to SU(6)'\times SU(2)'_{\theta} \to SU(3)'_C\times
SU(2)'_L\times SU(2)'_{\theta}\times U(1)'_X \times U(1)'_Z
\lb{3C} \ee
in the H-world, heavy Majorana neutrinos $N_a$ become singlets
of the subgroups $SU(3)_C\times SU(2)_L \times U(1)_X\times U(1)_Z$
and $SU(3)'_C\times SU(2)'_L\times U(1)'_X \times U(1)'_Z$,
according to Eq.~(\ref{9a}). Therefore, in our model \ct{1}, after
the breaking of $ SO(10)$ and $SU(6)'$ and below seesaw scale
($\mu < M_R=M'_R\sim 10^{10-15}$ GeV), when we have the symmetry
groups $G_{SM}$  and $G_{SM'}\times SU(2)'_{\theta}$, the heavy
Majorana neutrinos $N_a$ again can play the role of messengers
between O- and H-worlds.

Baryon $B$ and lepton $L$ numbers are not perfect quantum numbers.
They are directly related to the seesaw mechanism for light
neutrino masses. $B - L$ is generated in the decays of heavy
Majorana neutrinos, $N$, into leptons $l$ (or anti-leptons $\bar
l$) and the Higgs bosons $\phi$ (which are the standard Higgs
doublets):
\be
      N \to l\phi,\,\,\bar l\bar \phi.  \lb{4C} \ee
In this context, the three necessary Sakharov conditions are
realized in the following way:

1) $B-L$ and $L$ are violated by the heavy neutrino Majorana
masses.

2) The out-of-equilibrium condition is satisfied due to the
delayed decay(s) of the Majorana neutrinos, when the decay rate
$\Gamma(N)$ is smaller than the Hubble rate $H$: $\Gamma(N)< H$,
i.e. the life-time is larger than the age of the Universe at the
time when $N_a$ becomes non-relativistic.

3) CP-violation (C is trivially violated due to the chiral nature
of the fermion weak eigenstates) originates as a result of the
complex $lN\phi$ Yukawa couplings producing asymmetric decay
rates:
\be    \Gamma(N\to l\phi)\neq  \Gamma(N\to \bar l \bar{\phi}),
\lb{5C} \ee
so that leptons and anti-leptons are produced in different amounts
and the $B-L$ asymmetry is generated.

In the present model, the quantum numbers $B$ and $L$ are related to
the accidental global symmetries existing at the level of
renormalizable couplings, which can be explicitly broken by higher
order operators with the large mass scale M as a cutoff. In
particular, the  $D=5$ operator
\be {\cal O}_5 \sim \frac 1M(l\phi)^2 \qquad (\Delta L=2) \lb{6C}
\ee
yields the small Majorana masses for neutrinos according to the
seesaw mechanism, $m_{\nu}\sim v^2/M$, where $v$ is the Higgs VEV
given by Eq.~(\ref{1}).

As for the H-sector, the shadow neutrinos get masses via the
operator:
\be {\cal O'}_5 \sim \frac 1M(l'\phi')^2\qquad (\Delta L'=2),
\lb{7C} \ee
which yields the small Majorana masses for shadow neutrinos:
$m'_{\nu}\sim v'^2/M$, where $v'$ is the shadow Higgs VEV in
Eq.~(\ref{1}). However, there can exist also a mixed gauge invariant
operator:
\be {\cal O}_5^{mix} \sim \frac 1M(l\phi)(l'\phi')\qquad (\Delta
L=1,\,\,\,\Delta L'=1), \lb{8C} \ee
that gives rise to the mixing between the ordinary and shadow
neutrinos. All these operators can be induced by the same seesaw
mechanism.

Considering $n$-species ($n$-generations) of the heavy Majorana
neutrinos $N_a$ with the large mass terms $M_Ng_{ab}N_aN_b$, we
use $M=M_N$ as an overall mass scale. The matrix $g_{ab}$ of
dimensionless Yukawa-like constants ($a,b= 1,2, ...,n$) is taken
diagonal without lose of generality. Remembering that $N_a$ are
gauge singlets, playing the role of messengers between the
ordinary and shadow worlds, we assume that they would couple the
ordinary leptons $l_i = (\nu, e)_i$ and shadow leptons $l'_i =
(\nu', e')_i$ with similar rights:
\be     Y_{ia}l_iN_a\phi + Y'_{ia}l'_iN_a\phi'. \lb{9C} \ee
In the framework of the seesaw mechanism, we obtain the following
operators:
\be {\cal O}_5 = \frac{A_{ij}}{M}(l_i\phi)(l_j\phi), \qquad
 {\cal O'}_5 = \frac{A'_{ij}}{M}(l'_i\phi')(l'_j\phi'),\qquad
 {\cal O}_5^{mix} = \frac{D_{ij}}{M}(l_i\phi)(l_j'\phi'),
\lb{10C} \ee
with the following coupling constant matrices:
\be A = Yg^{-1}Y^T,\quad A'= Y'g^{-1}Y'^T,\quad D = Yg^{-1}Y'^T.
  \lb{11C} \ee
The Yukawa constant matrices obey the relation: $Y'= Y^*$, giving
$A' = A^*$ and $D = D^+$.

The interactions mediated by heavy neutrinos $N_a$ induce the
processes $l\phi\to \bar{l}\bar {\phi}$, etc. in the O-world (with
$\Delta L=2$), $l'\phi'\to \bar{l'}\bar{\phi'}$, etc. in the
H-world (with $\Delta L'=2)$, and the processes $l\phi\to
\bar{l'}\bar{\phi'}$, etc. with $\Delta L=1,\,\, \Delta L'=1$ that
transform O-particles into H-partners.

It is easy to see that all three conditions for baryogenesis
\ct{20} are naturally fulfilled:

1) $B-L$ violation is obvious: there are processes which,
conserving $B(B')$, violate $L(L')$ and thus both $B-L$ and
$B'-L'$.

2) CP violation in these processes is fulfilled  due to the complex
Yukawa matrices $Y$ and $Y'$. As a result, the cross-sections with
leptons and anti-leptons in the initial state are different from
each other. CP-asymmmetry emerges in processes with $\Delta L=1$,
as well as with $\Delta L=2$, due to the interference between the
tree-level and one-loop diagrams shown in Refs.~\ct{18}. The
diagrams relevant for $l\phi\to \bar{l'}\bar{\phi'}$ are shown in
Fig.~1. The diagrams responsible for CP-violation in $l\phi\to
\bar l\bar{\phi}$ and $\bar l\bar{\phi}\to l\phi$ are shown in
Fig.~2.

The direct calculation gives:
$$  \sigma(l\phi\to \bar l'\bar{\phi'}) -  \sigma(\bar
l\bar{\phi})\to l'\phi') = (-\Delta \sigma - \Delta \sigma')/2, $$
$$ \sigma(l\phi\to l'\phi') -  \sigma(\bar
l\bar{\phi})\to \bar l'\bar{\phi}') = (-\Delta \sigma + \Delta
\sigma')/2, $$ \be \sigma(l\phi\to \bar l\bar{\phi}) - \sigma(\bar
l\bar{\phi})\to l\phi) = \Delta \sigma  \lb{13C} \ee
with
\be \Delta \sigma= \frac{3JS}{32\pi^2M^4},\qquad \Delta \sigma' =
\frac{3J'S}{32\pi^2M^4}, \lb{14C} \ee
where $S$ is the c.m. energy square, and $J$ and $J'$ are the
CP-violation parameters:
$$ J={\rm{ImTr}}[g^{-1}(Y^+Y)^*g^{-1}({Y'}^+Y')g^{-2}(Y^+Y)], $$ \be
J'={\rm{ImTr}}[g^{-1}({Y'}^+Y')^*g^{-1}(Y^+Y)g^{-2}({Y'}^+Y')] .
 \lb{15C} \ee

3) All $\Delta L=1$ processes $l\phi\to l'\phi'$ and $\Delta L=2$
ones $l\phi\to \bar l\bar{\phi}$, $ll\to\phi \phi$ stay out of
equilibrium.

Many details of this model can be extracted from
Refs.~\ct{18}.

The reheating temperature $T_R$, at which the inflaton decay and
entropy production of the Universe are over and the relativistic
particles are dominated, plays a crucial role in cosmological
evolution. In our model and in \ct{18,19} $T'<T$ after the
postinflationary reheating, i.e. the shadow sector is cooler than
the ordinary one and almost "empty".

The assumption $M_N > T_R$ forbids the thermal production of heavy
neutrinos, and the usual leptogenesis mechanism \ct{22} via decays
$N\to l\phi$ does not work. However, a net $B-L$ may emerge in the
Universe due to the CP-violation in the processes $l\phi\to
l'\phi'$.

We see that in our  model the baryon asymmetry is generated not
only in the O-sector, but also in the H-sector. These two sectors
are not identical, but they have similar CP-violating properties
due to the complex coupling constants of scattering processes.
These processes are most effective at temperatures $T\sim T_R$,
although they stay out of equilibrium. Finally, at the relevant
epoch, the O-observer detects: (a) the loss of entropy in the
O-world due to the leakage of O-particles to the H-world; (b) the
leakage of leptons $l$ from O-sector to the H-sector with
different rates than anti-leptons $\bar l$, and as a result,
non-zero $B-L$ in the Universe. In parallel, the H-observer detects:
(a') entropy production in the H-world; (b') leptons $l'$ and
anti-leptons $\bar l'$ production with different rates, and
non-zero $B'-L'$.

Here it is necessary to comment that the baryon asymmetries in the
O- and H-sectors are not equal: by reason that H-world is colder
than O-world, the H-baryon asymmetry can be about one order of
magnitude bigger than the O-baryon asymmetry. This could explain the
difference between the O- and H-worlds.

The present work opens the possibility to specify a grand
unification group $E_6$ from cosmology.

\section*{Acknowledgements} We are grateful to Masud Chaichian for useful discussions. The
support of the Academy of Finland under the projects no. 121720
and 127626 is acknowledged. L.V.L. thanks RFBR grant
09-02-08215-3. C. R. Das gratefully acknowledges a scholarship
from Funda\c{c}\~{a}o para a Ci\^{e}ncia e Tecnologia ref.
SFRH/BPD/41091/2007.

\newpage

\begin{figure}[h]
\par
\begin{center}
\includegraphics*[width=80mm]{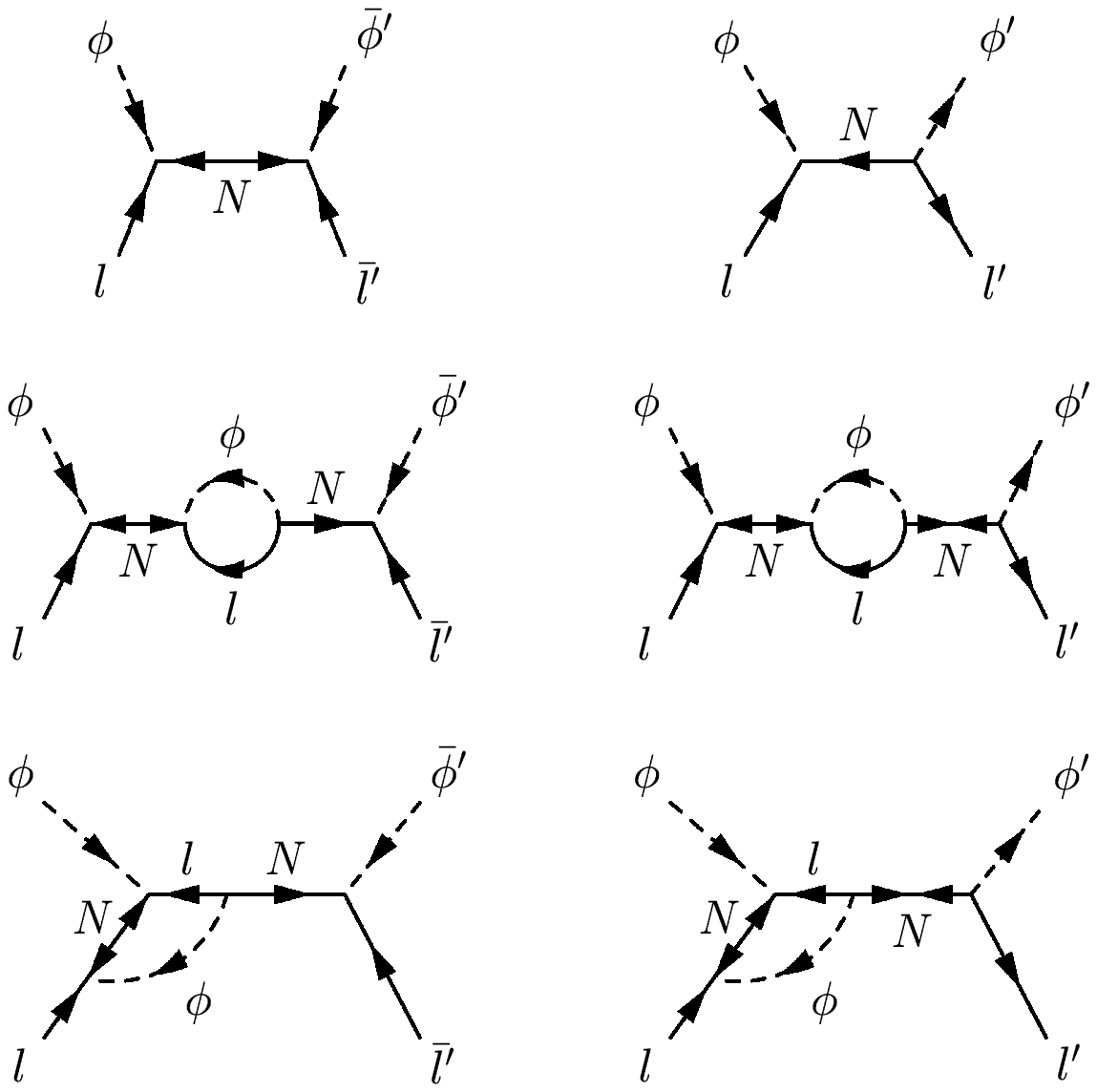}
\end{center}
\caption{Tree-level and one-loop diagrams contributing to the $CP$%
-asymmetries in the processes $l\phi \rightarrow \bar{l}^{\prime
}\bar{\phi}^{\prime }$ (left column) and $l\phi \rightarrow
l^{\prime }\phi ^{\prime }$ (right column).  Not all the vertex
corrections in the above Feyman diagrams are depicted. }
\label{fig1}
\end{figure}

\begin{figure}[h]
\begin{center}
\includegraphics*[width=80mm]{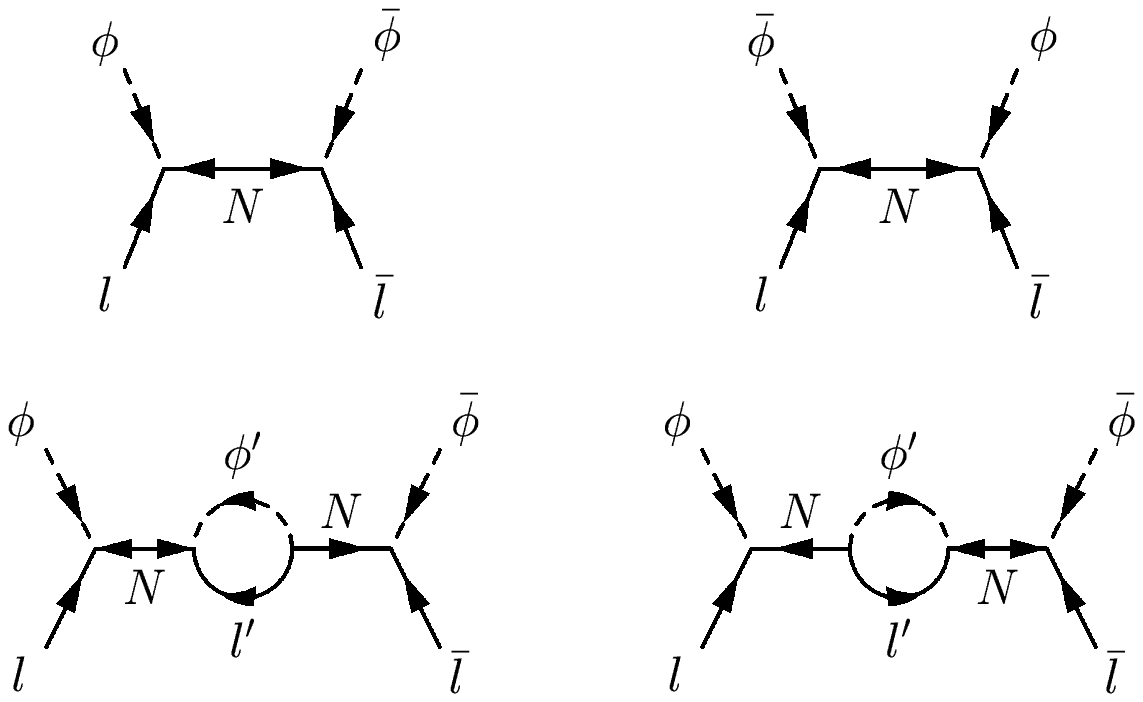}
\end{center}
\caption{Tree-level and one-loop diagrams contributing to the
CP-asymmetry in the process $l\phi \rightarrow \bar{l}\bar{\phi}$.
The vertex diagrams corrections are not depicted.} \label{fig2}
\end{figure}


\begin{thebibliography}{99}

\bibitem{1}
C.R.~Das, L.V.~Laperashvili, A.~Tureanu, Eur.Phys.J.C {\bf 66}
(2010) 307; arXiv:0902.4874; AIP Conf.Proc. {\bf 1241} (2010) 639;
arXiv:0910.1669.
\bibitem{1a}
P. Q.~Hung, Nucl. Phys. B {\bf 747} (2006) 55; J. Phys. A {\bf 40}
(2007) 6871; P.Q.~Hung, P.~Mosconi, hep-ph/0611001; M.~Adibzadeh
and P. Q.~Hung, Nucl. Phys. B {\bf 804} (2008) 223; H.~Goldberg,
Phys. Lett. B {\bf 492} (2000) 153; C. R.~Das and L.
V.~Laperashvili, Int. J. Mod. Phys. A {\bf 23} (2008) 1863; Phys.
Atom. Nucl. {\bf 72} (2009) 377.
\bibitem{2}
Particle Data Group, C.~Amster {\it et al.,} Phys. Lett. B {\bf
667} (2008) 1.
\bibitem{3}
A.~Riees et al.,  Astrophys. J. Suppl. {\bf 183} (2009) 109;
arXiv:0905.0697; W.L.~Freedman et al., Astrophys. J. {\bf 704}
(2009) 1036; arXiv:0907.4524; R.~Kessler at al., arXiv:0908.4274.
\bibitem{6}
J. H.~Schwarz, Phys. Rept. {\bf 89} (1982) 223; M. B.~Green, Surv.
High. En. Phys. {\bf 3} (1984) 127; M. B.~Green and J. H.~Schwarz,
Phys. Lett. B {\bf 149} (1984) 117; ibid., B {\bf 151} (1985) 21.
\bibitem {7}
D. J.~Gross, J. A.~Harvey, E.~Martinec and R.~Rohm, Phys. Rev.
Lett. {\bf 54} (1985) 502; Nucl. Phys. B {\bf 256} (1985) 253;
ibid., B {\bf 267} (1986) 75; P.~Candelas, G. T.~Horowitz,
A.~Strominger and E.~Witten, Nucl. Phys. B {\bf 258} (1985) 46.
\bibitem{8}    
M. B.~Green, J. H.~Schwarz and E.~Witten, {\it Superstring theory}
(Cambridge University Press, Cambridge, 1988).
\bibitem{9}
K. Nishijima and M. H. Saffouri, Phys. Rev. Lett. {\bf 14} (1965)
205; L. B.~Okun and I. Ya.~Pomeranchuk, JETP Lett. {\bf 1} (1965)
167; Phys. Lett. {\bf 16} (1965) 338; E. W.~Kolb, D.~Seckel, M. S.
Turner, Nature {\bf 314} (1985) 415; Fermilab-Pub-85/16-A,
Jan.1985.
\bibitem{9a}
L. B.~Okun, Phys. Usp. {\bf 50} (2007) 380, hep-ph/0606202; S.
I.~Blinnikov, {\it Notes on Hidden Mirror World}, arXiv:0904.3609.
\bibitem{10}
T. D.~Lee and C. N.~Yang, Phys. Rev. {\bf 104} (1956) 254.\\
I. Yu.~Kobzarev, L. B.~Okun and I. Ya.~Pomeranchuk, Yad. Fiz. {\bf
3} (1966) 1154 [Sov. J. Nucl. Phys. {\bf 3} (1966) 837].
\bibitem{11}
P.~Athron, F.~King, D. J.~Miller, S.~Moretti and R.~Nevzorov,
Phys. Rev. D {\bf 80} (2009) 035009; arXiv:0904.2169;
arXiv:0901.1192.
\bibitem{12}
Z.~Berezhiani, A.~Dolgov and R. N.~Mohapatra, Phys. Lett. B {\bf
375} (1996) 26; Z.~Berezhiani and R. N.~Mohapatra, Phys. Rev. D
{\bf 52} (1995) 6607; Z.~Berezhiani, {\it Through the
looking-glass: Alice's adventures in mirror world}, in: Ian Kogan
Memorial Collection ``From Fields to Strings: Circumnavigating
Theoretical Physics'', Eds. M.~Shifman et al., World Scientific,
Singapore, Vol.~3, pp. 2147-2195, 2005; Acta Phys. Polon. B {\bf
27} (1996) 1503; Int. J. Mod. Phys. A {\bf 19} (2004) 3775;
Z.~Berezhiani, D.~Comelli and F. L.~Villante, Phys. Lett. B {\bf
503} (2001) 362; Z. Berezhiani, P. Ciarcelluti, D. Comelli and F.
L. Villante, Int. J. Mod. Phys. D {\bf 14} (2005) 107;
Z.~Berezhiani, L.~Pilo, N.~Rossi, arXiv:0902.0146.
\bibitem{14}
R.~Slansky, Phys. Rept. {\bf 79} (1981) 1.
\bibitem{15}
Taichiro Kugo, Joe Sato, Prog. Theor. Phys. {\bf 91} (1994) 1217.
\bibitem{16}
L. B.~Okun, JETP Lett. {\bf 31} (1980) 144; Pisma Zh. Eksp. Teor.
Fiz.{ \bf 31} (1979) 156; Nucl. Phys. B {\bf 173} (1980) 1.
\bibitem{17}
Z.~Berezhiani, D.~Comelli and N.~Tetradis, Phys. Lett. B {\bf 431}
(1998) 286.
\bibitem{18}
L. Bento and Z. Berezhiani, Phys. Rev. Lett. {\bf 87} 231304
(2001);  Fortsch. Phys. {\bf 50} (2002) 489; Z. Berezhiani, in:
AIP Conf. Proc. {\bf 878} (2006) 195;  Eur. Phys. J. ST {\bf 163}
(2008) 271.
\bibitem{19}
Z.~Berezhiani, L.~ Kaufmann, P.~Panci, N.~Rossi, A.~Rubbia,
A.~Sakharov, {\it Strongly interacting mirror dark matter},
CERN-PH-TH-2008-108, May 2008.
\bibitem{20}
A.D.~Sakharov, Pisma Zh. Eksp. Teor. Fiz. {\bf 5} (1967) 32.\\
V.A.~Kuzmin, V.A.~Rubakov and M.E.~Shaposhnikov, Phys. Lett. B
{\bf 155} (1985) 36.
\bibitem{21}
A.D.~Dolgov, AIP Conf.Proc. {\bf 1116} (2009) 155;
arXiv:0901.2100.
\bibitem{22}
Z.~Berezhiani, S.~Cassisi, P.~Ciarcelluti and A.~Pietrinferni,
Astropart. Phys. {\bf 24} (2006) 495.

\end{thebibliography}
\end{document}